# Upper critical dimension of the KPZ equation

Moshe Schwartz

School of Physics and Astronomy,
Raymond and Beverly Faculty of Exact Sciences,
Tel Aviv Univrersity, 69978 Tel Aviv, Israel

and

Ehud Perlsman

Simtat Hagina 11, Ramat Hasharon 47207,
Israel

Abstract

Numerical results for the Directed Polymer model in 1+4 dimensions in various types of disorder are presented. The results are obtained for system size considerably larger than that considered previously. For the extreme strong disorder case (Min-Max system), associated with the Directed Percolation model, the expected value of the meandering exponent, $\zeta$ = 0.5 is clearly revealed, with very week finite size effects. For the week disorder case, associated with the KPZ equation, finite size effects are stronger, but the value of $\zeta$ is clearly seen in the vicinity of 0.57. In systems with "strong disorder" it is expected that the system will cross over sharply from Min-Max behavior at short chains to weak disorder behavior at long chains. This is indeed what we find. These results indicate that 1+4 is not the Upper Critical Dimension (UCD) in the week disorder case, and thus 4+1 does not seem to be the upper critical dimension for the KPZ equation.

The importance of the KPZ equation [1] goes far beyond providing a description of a surface growing under ballistic deposition. Besides being mathematically equivalent to a number of interesting physical problems, it also provides a relatively simple prototype of non-linear stochastic field equations, which are so abundant in condensed matter physics.

The equation for the height of the surface, $h$, at a point $\mathbf{r}$ and time $t$ is given by

$$\frac{\partial h}{\partial t} = \nu \nabla^2 h + g(\nabla h)^2 + \eta, \tag{1}$$

where $\eta(\mathbf{r},t)$ is a noise term such that

$$\langle \eta(\mathbf{r},t) \rangle = 0 \text{ and } \langle \eta(\mathbf{r},t)\eta(\mathbf{r}',t') \rangle = 2D\delta(\mathbf{r}-\mathbf{r}')\delta(t-t'). \tag{2}$$

The steady state correlation

$$\Phi(\Delta \mathbf{r}, \Delta t) = \lim_{t \to \infty} \langle [h(\mathbf{r}+\Delta \mathbf{r}, t+\Delta t) - h(\mathbf{r},t)]^2 \rangle, \tag{3}$$

is characterized by two exponents, the roughness exponent $\alpha$ and the growth exponent $\beta$ defined by the relations

$$\Phi(\Delta \mathbf{r}, 0) \propto |\Delta \mathbf{r}|^{2\alpha} \text{ and } \Phi(0, \Delta t) = |\Delta t|^{\beta}. \tag{4}$$

(A third exponent used in the description of the system is the dynamic exponent $z$ but it is not independent of the first two and is given by

$$z = \frac{\alpha}{\beta}.)$$

It is well known that for dimensions $d \leq 2$ the surface is always rough ($\alpha > 0$ for $d < 2$ and logarithmic roughness for $d = 2$). Above two dimensions the picture changes and two phases exist depending on the dimensionless strength of the non-linearity. In the weak coupling phase the surface is flat and the steady state correlation can be obtained perturbatively in the non-linear coupling. In the strong coupling regime, the surface is rough and perturbation theory cannot be used. There is, however, a long lasting controversy concerning the existence of an upper critical dimension (UCD). Namely, does there exist a dimension above which, regardless of the strength of the non-linearity, the surface remains flat? Historically, the question seems to stem from a paper by Cook and Derrida [2]. They considered the directed polymer problem, which by the Hopf Cole transformation is equivalent to the KPZ system. Expanding in the inverse dimension of the system, they concluded that that system has an upper critical dimension. The idea that a UCD exits gained support following the introduction of the self consistent expansion (SCE), which was used successfully in the treatment of KPZ and its many variants [3-9], followed by the use of mode coupling (MC) [10,11] which is different but produces equations not entirely dissimilar to SCE. Both schemes lead to integral equations that [10,12] have a strong coupling solution only for dimensions $d \leq d_c$, where $d_c$ is between three and four [10,12]. This was followed up by independent analytic studies that claim to have shown that indeed a UCD exists and is bound from above by four [13] or even were led to the conjecture that it is exactly four [14].

On the other hand, there is accumulated evidence that an upper critical dimension does not exist. Real space RG treatment of the directed polymer problem [15] and independent Real space RG on a discrete KPZ system [16], while yielding similar results to one another, find no UCD. More compelling are the numerical results by Ala-Nissila et. al. [17,18] and Kim [19]. Reference [17] presents simulations on the RSOS system and finds no upper critical dimension up to $d = 7$, while ref. [18], which is a response to the paper of Lässig and Kinzelbach [13], concentrates mainly on $d = 4$. The paper of Kim [19], which is also a response to [13] studies the directed polymer system at a finite temperature in $4$ (transverse) dimensions as a function of temperature. It shows a transition as a function of temperature from a weak coupling behavior at high temperatures to a strong coupling behavior at low temperatures. Thus, while not answering the general question of whether a UCD exists or not, he concludes that an upper critical dimension does not exist for $d \leq 4$, in contradiction to the claims in refs. [13,14].

Clearly the real space RG could be easily dismissed by claiming that they represent only approximations that cannot be trusted when the differences between the weak and strong coupling exponents are slight. As for the numerical results, Lässig and Kinzelbach [20] claim that those systems are too small to provide reliable results.

Here we consider in $d=4$ (transverse) dimensions the directed polymer problem at zero temperature [21] for a system which is about eight times larger than that considered by Kim. We obtain the meandering exponent, $\zeta$, defined by

$$\langle \mathbf{r}^2(t) \rangle \propto t^{2\zeta} \text{ for large } t,  \tag{5}$$

where the $\mathbf{r}(t)$ is the $d$ projection of the $d+1$ position vector $(\mathbf{r}(t),t)$, which describes the directed polymer. First, we study the meandering of the directed polymer in a "weak disorder"[22] background. Namely, the independent random values attached to each bond are governed by a Gaussian distribution that prefers naturally values which are close to the mean. We obtain $\zeta$ slightly higher than 0.57, which is very close to that of Kim [19] and to that of Alaa-Nissila et. al. [17,18], when translated from the surface roughening language to the directed polymer language. Because the value of $\zeta$ in the weak coupling regime is 0.5, not very far from 0.57, it is important to verify that (a) our procedure is able to distinguish between the two and (b) that the value obtained is not a result of some inherent inaccuracy in that procedure. We do that by considering two additional systems defined on the same lattice for which we know exactly what to expect as will be detailed in the following. We consider first the Min-Max system [23,24]. This system is defined by the following procedure: First a subset of directed polymer configurations, $\Omega_1$ is chosen, such that the maximal energy associated with a link belonging to that configuration is the same for all configurations belonging to $\Omega_1$ yet it is lower from the energy of the maximal link for any configuration, $C$ such that $C \notin \Omega_1$. Next a subset of $\Omega_1$, $\Omega_2$, is chosen, such that the second largest energy associated with a link belonging to that configuration is the same for all configurations belonging to $\Omega_2$ yet it is lower from the energy of the second largest link for any configuration, $C$ such that $C \in \Omega_1 - \Omega_2$. The procedure continues until a single configuration is chosen. It is well known [25] that the Min-Max problem is associated with the directed percolation problem. In fact the meandering exponent of the Min-Max optimal configuration is given by [25]

$$\zeta = \frac{\nu_\perp}{\nu_\parallel}, \tag{6}$$

where $\nu_\perp$ and $\nu_\parallel$ are the transverse and longitudinal correlation length exponents of directed percolation clusters respectively. For the directed percolation problem it is known that four is the upper critical (transverse) dimension and that for $d \geq 4$ the exponents $\nu_\perp$ and $\nu_\parallel$ attain their mean field values ½ and 1 respectively. Consequently the meandering exponent for the Min-Max optimal configuration in four (transverse) dimensions should be just 0.5.

The next system we consider, is a " strong disorder" directed polymer system. To generate "strong disorder" [22] the values of the (dimensionless) independent energies, $x$ ,associated with each bond are taken from a wide distribution,

$$P_k(x) = \begin{cases} 1/kx & \text{for } 1 \leq x \leq \exp(k) \\ 0 & \text{otherwise} \end{cases}. \tag{7}$$

The width of the distribution increases with $k$ and for a given finite length , $T$ , of a directed polymer in a random environment governed by the above distribution it is possible to find $k$ large enough that single maximal links dominate the sum of links in directed polymer configurations [ 23,24]. It is obvious, on the other hand, that for given $k$ it is possible to find large enough $T$ such that a single term cannot dominate the sum. Thus the conclusion is that for fixed $k$ the behavior of the directed polymer will cross over as a function of $T$ from Min-Max behavior to regular directed polymer behavior. In the graph below we present the local meandering exponent, defined as

$$\zeta(t) \equiv \frac{1}{2}\log_2\{\langle \mathbf{r}^2(t)\rangle / \langle \mathbf{r}^2(t/2)\rangle\}, \tag{8}$$

for the three systems described above. The local meandering exponent for the ordinary directed polymer system with the Gaussian distribution of bond strength,

$$P(x) = \frac{1}{\sqrt{2\pi}} \exp(-\frac{1}{2}x^2) \qquad (9)$$

increases smoothly from 0.5 at small $t$ and rises above 0.57 in the vicinity of $t=200$, where it seems to flatten off. The local meandering exponent for the Min-Max system sticks to 0.5 for all $t$ and the "strong disorder" local exponent, for a system described by the distribution $P_2(x)$ (equation(7)), clearly crosses over between Min-Max and the ordinary directed polymer local exponents.

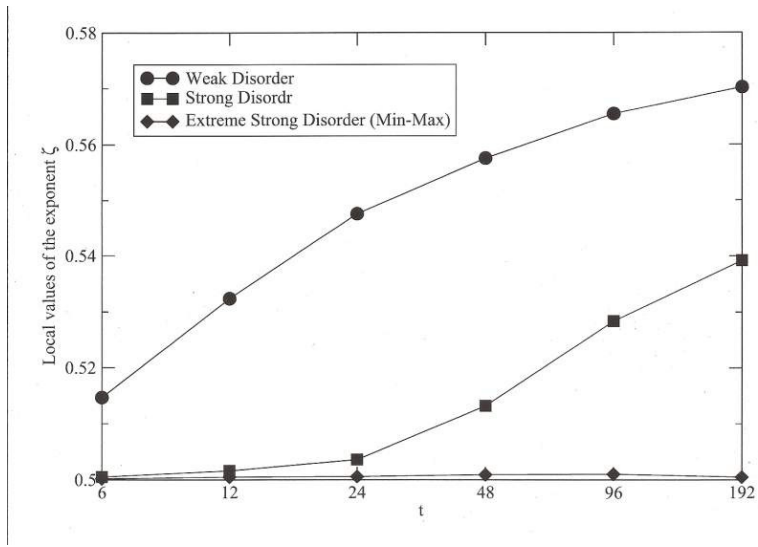

Figure 1

The local of the exponent $\zeta$ for the weak disorder (regular KPZ), strong disorder and the Min-Max systems.

The seemingly obvious conclusion from the above is that the (transverse) dimension four lies below the upper critical dimension for the directed polymer problem and consequently this is also the case for the KPZ system. Is that the only possible conclusion? It seems that there is a remote possibility that the claims that $d_c \leq 4$ may be still consistent with our finding if we drop the assumption that the continuous directed polymer problem and KPZ system are equivalent to their discrete counterparts. In fact, the suspicion that there may be serious differences between discrete KPZ and its continuous counterpart above one dimension was suggested in the past [26]. Thus it seems that the final resolution of the controversy hinges on the question whether these differences are severe enough to break the assumed universality.